# THE EVOLUTION OF FTK, A REAL-TIME TRACKER FOR HADRON COLLIDER EXPERIMENTS

A. Annovi[*], M. Beretta, and P. Laurelli
*INFN Frascati*

E. Bossini, V. Cavasinni, F. Crescioli, M. Dell'Orso, P. Giannetti,
M. Piendibene, G. Punzi, F. Sarri, I. Vivarelli, G. Volpi, and L. Sartori
*Univ. and INFN of Pisa*

A. Boveia, E. Brubaker, F. Canelli, M. Dunford, A. Kapliy, Y.K. Kim,
C. Melachrinos, M. Shochet, and J. Tuggle
*Univ. of Chicago*

H. DeBerg, A. McCarn, and M. Neubauer
*Univ. of Illinois at Urbana-Champaign*

M. Franklin and C. Mills
*Harvard Univ.*

N. Kimura and K. Yorita
*Waseda University*

J. Proudfoot and J. Zhang
*Argonne National Lab*

R. Tripiccione
*Univ. and INFN Ferrara*

We describe the architecture evolution of the highly-parallel dedicated processor FTK, which is driven by the simulation of LHC events at high luminosity ($10^{34}$ cm$^{-2}$ s$^{-1}$). FTK is able to provide precise on-line track reconstruction for future hadronic collider experiments. The processor, organized in a two-tiered pipelined architecture, execute very fast algorithms based on the use of a large bank of pre-stored patterns of trajectory points (first tier) in combination with full resolution track fitting to refine pattern recognition and to determine off-line quality track parameters. We describe here how the high luminosity simulation results have produced a new organization of the hardware inside the FTK processor core.

---

[*] email: alberto.annovi@lnf.infn.it

## 1. Introduction

Hadron collider experiments require large online computing power to reach the enormous rejection factor necessary to select events to be written on tape. One of the most demanding tasks is usually the track reconstruction from measured points (detector *hits*) on particle trajectories. In this paper we propose the use of highly-parallel dedicated processors to efficiently execute two fast track finding algorithms[1]. The algorithms are based on the idea of a large bank of pre-calculated hit patterns to be compared with the event ("pattern matching")[2].

We propose FTK for the ATLAS experiment[3] and we simulate its performances for ATLAS. However FTK has a general structure that can easily fit with different applications. FTK is a scalable processor. The FTK could be easily integrated in modern data acquisition systems. It needs to "spy" a copy of the tracker data to be processed. It will provide as output reconstructed track parameters that can be stored in a buffer memory ready to be accessed by CPUs. A description of FTK connected to the ATLAS Trigger and DAQ is available[3]. In the rest of this paper we will describe the new FTK architecture shown in fig. 1 and the reasons that lead to this development from the original design[3].

## 2. FTK architecture overview

The first FTK requirement is to receive all the data read out from the silicon tracker. In order to sustain very high event rates, it is necessary to organize FTK as a set of independent engines (input bandwidth of 4 Gbit/sec each), each one working on a different sector of the silicon tracker. Let us imagine dividing the detector into azimuthal ($\phi$) sectors. At high luminosity we plan to use 8 $\phi$ sectors that we call "regions", as originally proposed in Ref. 3. Eight processors equal to the one shown in fig. 1 will work in parallel to reconstruct the whole ATLAS event. In order to further increase parallelization and data throughput for high luminosity, within each region the processor is organized into 6 $\eta$-$\phi$ towers (as shown in fig.1) for a total of 48 independent engines.

Within each $\eta$-$\phi$ tower an FTK slice performs the most CPU consuming part of the pattern recognition. It splits the problem of finding tracks inside the whole detector into many simpler problems of finding tracks inside detector "slices" called roads. The FTK slice consists of a Road Finder, a Data Organizer (DO) and a Track Fitter (TF). The Road Finder looks for low-resolution track candidates in the whole detector. The resolution must be low enough to keep costs under control and to make the problem solvable at high rate. Widths too small or too large would require intolerably high performance of either the Road Finder or the Data Organizer and Track Fitter. The roads, however, have to be narrow enough to hold only a few hit ambiguities, if any. These ambiguities can be solved by sequentially fitting the residual hit combinations and choosing the best fit.

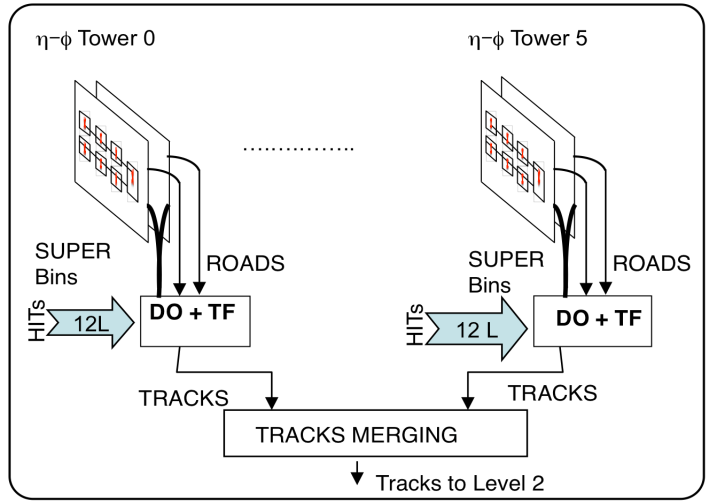

Fig. 1: The pipeline of 12 AM boards broken into 6 independent engines each with its own DO and TF working in parallel.

The Data Organizer is a high-speed data traffic node between the detector, the Road Finder, and the Track Fitter. The Data Organizer performs the following operations: (a) receives full resolution detector hits in any order, (b) buffers them in an internal database, (c) sends low-resolution hits called super bins to the Road Finder (the super bins are obtained by logically ORing a number of adjacent detector bins or channels), (d) receives roads from the Road Finder and fetches from the internal data base all the detector hits contained in the roads, (e) sends each road with its set of full resolution hits to the Track Fitter. The new organization into η-φ towers allows also parallelization of the Data Organizer and Track Fitter functions by breaking them into 6 independent engines working in parallel on 6 subsets of the region's data.

## 3. An additional pattern matching step

The original architecture with pipelined AM has been tested by simulating WH events hidden in the pile-up of the LHC high luminosity environment ($10^{34}$ cm$^{-2}$ s$^{-1}$) with the FTKsim package described in Ref. 4. We have chosen initially half the road size used in Ref. 4: super bins in the r–$\varphi$ plane of 2.5mm for Pixels and 5mm for SCT detectors; both extend in $z$ for the length of a full module. This choice is what we call "the default road size". Even if the quality of FTK tracking has been demonstrated to be extremely good for high Pt events, as found for soft B-physics in Ref. 4, the extremely high average number of roads in the pipeline and content of multi hits inside roads has constrained us to reorganize the FTK architecture.

We have studied different architecture improvements to increase the power of the pattern-matching step at an instantaneous luminosity of $10^{34}$ cm$^{-2}$ s$^{-1}$, so that the workload of the Data Organizer and the Track Fitter boards can be sharply reduced. Here we describe the option that uses the Tree Search Processor (TSP) to provide finer pattern matching resolution thus reducing the number of roads to be handled by the Data Organizer and the number of fits to be performed by the Track Fitter.

If the roads produced by the Associative Memory are too wide (we call them "fat roads") to adequately scale down the combinatorial problem, an intermediate step of pattern recognition at higher resolution can be performed by the Tree Search Processor (TSP) before the final track fit. It is designed for this purpose as it takes fat roads in input and it outputs thin roads. It is dedicated parallel and pipelined hardware that implements a binary search strategy inside the pattern bank. High density commercial RAMs store the bank; the machines are easily packed into FPGA devices. A complete description of the implementation of the processor is available in Ref. 5.

## 4. Results of the simulation to check the new FTK

We have repeated the FTK simulation using a large bank of 108 Million patterns (easily implemented in the TSP using commercial RAMs), and a not yet optimized choice of thin roads. The SCT super bin size has been reduced a factor 2 in the r–$\varphi$ plane and the pixel super bin size has been divided a factor 4 in the z direction, both compared to the default road. The average number of fits per event is reduced to 400000 and will be partitioned into 6 engines working in parallel, so the number of fits to be executed per ns is 5, a number that could be supported today by the existing technology[6]. The DO has more than 10 ns on average to process a road. Pushing to better resolution in the TSP pattern matching, or better road shape, the available time can be increased.

The number of hits to be loaded on average in the AM in the whole region is roughly 1000 per event per input bus. No more than a third will be necessary for each η–ϕ tower, so no more that 350 hits will go on average to each DO and each AM board. The global input bandwidth for an event rate of 75 kHz will be 25 MHz that is compatible even with the input clock frequency of the old AM chip[7]. In conclusion the new architecture and the AM board prototype we have today are both compatible with an FTK that could work up to $10^{34}$ cm$^{-2}$ s$^{-1}$.

## 5. Conclusions

The simulation of WH events at the instantaneous luminosity $10^{34}$ cm$^{-2}$ s$^{-1}$ has produced changes to the original FTK architecture to increase the parallelism of the Data Organizer and Track Fitter functions. Studies at the SLHC luminosity of $3*10^{34}$ cm$^{-2}$ s$^{-1}$ are currently in progress. The use of the TSP processor as an expansion of the AM allows an early use of FTK even with the old chip developed for CDF[7]. Future evolution of the system to sustain $3\times10^{34}$ cm$^{-2}$ s$^{-1}$ will be strongly based on an AM chip upgrade whose R&D is already ongoing[8]. The described processors can find tracks at the ATLAS L1 output event rate of 75 kHz. They can provide event reduction using tracks in trigger applications. Hits of track candidates, with P$_t$ above a threshold of 1 GeV and with impact parameters compatible with *b* quark decay, or with a tau-jet topology compatible with hadronic tau decays, can be filtered among a huge number of other hits. The ambitious goal of trigger selection of *b* and hadronic tau decays at the future hadron colliders can benefit from our architecture.